\documentclass[]{spie}  %>>> use for US letter paper
\usepackage{amsmath,amsfonts,amssymb}
\usepackage{graphicx}
\usepackage[colorlinks=true, allcolors=blue]{hyperref}
\usepackage{siunitx}

\newcommand{\appropto}{\mathrel{\vcenter{
  \offinterlineskip\halign{\hfil$##$\cr
\propto\cr\noalign{\kern2pt}\sim\cr\noalign{\kern-2pt}}}}}

\title{Progress toward a demonstration of high contrast imaging at ultraviolet wavelengths}

\author[a]{Kyle Van Gorkom}
\author[a]{Ramya M.\ Anche}
\author[b]{Christopher B.\ Mendillo}
\author[c]{Jessica Gersh-Range}
\author[a,d]{G.C. Hathaway}
\author[a]{Saraswathi Kalyani Subramanian}
\author[a]{Justin Hom}
\author[e]{Tyler D.\ Robinson}
\author[f]{Mamadou N'Diaye}
\author[g]{Nikole K.\ Lewis}
\author[h]{Bruce Macintosh}
\author[a]{Ewan S.\ Douglas}
\affil[a]{Steward Observatory, University of Arizona, 933 N Cherry Avenue, Tucson, Arizona, 85721, USA}
\affil[b]{Lowell Center for Space Science and Technology, University of Massachusetts Lowell, 600 Suffolk St. Suite 315, Lowell, MA 01854, USA}
\affil[c]{DM Telescopes LLC, Raleigh, NC, USA}
\affil[d]{James C. Wyant College of Optical Sciences, University of Arizona, Tucson, Arizona, USA}
\affil[e]{Lunar \& Planetary Laboratory, University of Arizona, 1629 E University Blvd, Tucson, AZ 85721, USA }
\affil[f]{Université C{\^o}te d’Azur, Observatoire de la C{\^o}te d’Azur, CNRS, Laboratoire Lagrange, Nice, France}
\affil[g]{Department of Astronomy, Cornell University, Space Sciences Bldg, 404, 122 Sciences Dr, Ithaca, NY 14850, USA}
\affil[h]{UC Observatories, Astronomy \& Astrophysics Department, University of California Santa Cruz, CA 95064, USA}
\authorinfo{Send correspondence to K.V.G., kvangorkom@arizona.edu}

\begin{document} 
\maketitle

\begin{abstract}
NASA's Habitable Worlds Observatory (HWO) aims to achieve starlight suppression to the $10^{-10}$ level for the detection and spectral characterization of Earth-like exoplanets. Broadband ozone absorption features are key biosignatures that appear in the \SI{200}-\SI{400}{\nano \meter} near-ultraviolet (UV) regime. Extending coronagraphy from visible wavelengths to the UV, however, brings with it a number of challenges, including tighter requirements on wavefront sensing and control, optical surface quality, scattered light, and polarization aberrations, among other things. We aim to partially quantify and address these challenges with a combination of modeling, high-resolution metrology to the scales required for UV coronagraphy, and---ultimately---a demonstration of UV coronagraphy on the Space Coronagraph Optical Bench (SCoOB) vacuum testbed. In these proceedings, we provide a status update on our modeling and contrast budgeting efforts, characterization efforts to understand performance limitations set by key optical components, and our plans to move toward a demonstration of UV coronagraphy.
\end{abstract}

% Include a list of keywords after the abstract 
\keywords{high contrast imaging, ultraviolet, coronagraphy, exoplanets}

\section{INTRODUCTION}\label{sec:intro}

The presence of ozone (O$_3$) in an exoplanet atmosphere is a significant biosignature in the search for an Earth-like exoplanet, one of the key science goals of the Habitable Worlds Observatory (HWO). O$_3$ absorption is a dramatic spectral signal that begins shortward of 350 nm and requires the development of near ultraviolet (NUV) high contrast imaging (HCI) capabilities to the $10^{-10}$ contrast levels. In spite of the strong science case for exoplanet observations in the NUV, relatively little effort has focused on pushing technology development beyond the existing capabilities of the Hubble Space Telescope Imaging Spectrograph (STIS) instrument which cannot combine filters with coronagraphy and thus operates at low contrast across a single 200nm –1030 nm bandpass. 

We aim to address this technology gap with a combination of end-to-end coronagraph simulations, the development of analytic tools to identify challenges and guide design trades, metrology of key optical components to high spatial resolution and with direct measurements in the NUV, and ultimately a demonstration of NUV coronagraphy in a vacuum testbed environment.

\section{CONTRAST CHALLENGES IN THE NUV}

A number of physical effects that could pose challenges to the development of a $10^{-10}$ contrast capable NUV coronagraph have been previously reported\cite{tuttle2024_hwo,ctr_2024,vangorkom_2025} and include imperfections in NUV optical coatings, polarization aberrations, contamination and scattered light control, and tighter requirements on the performance of wavefront sensing and control.

The wavelength-scaling of a subset of these terms---most notably chromatic residuals (due to the Talbot effect) after electric field conjugation (EFC)\cite{efc}, scattering, beamwalk (beam shear on optical surfaces due to line-of-sight jitter), and deformable mirror quantization and actuator noise---are highlighted in Van Gorkom et al.~2025\cite{vangorkom_2025}. At a separation that scales with resolution (i.e., in units of $\lambda/D$), the contrast limit set by these terms tends to go as $\lambda^{-2}$ or $\lambda^{-4}$, which suggests that a NUV coronagraph may be challenging to build to the contrast levels required for HWO. The inner working angle (IWA) required to detect and characterize an Earth-like exoplanet, however, suggests that a NUV coronagraph may have a relaxed IWA requirement (in $\lambda/D$ units) that may significantly ease the challenges of UV coronagraphy.

At a fixed angular separation on-sky (e.g. 9$\lambda/D$ at $\lambda=200$ nm and 3$\lambda_0/D$ at $\lambda_0=600$ nm), the contrast scaling appears to be more favorable. The ratio of the contrast at wavelength $\lambda$ compared to $\lambda_0$ from first-order phase aberrations on out-of-pupil-optics, scattering, and beam walk show an identical wavelength dependence\cite{vangorkom_whitepaper_2025}
\begin{equation}\label{eqn:contrast_scaling_general}
    C \propto \left( \frac{\alpha_n}{\alpha_{n,0}} \right)^2 \left( \frac{\lambda_0}{\lambda} \right)^4,
\end{equation}
where $\alpha_n$ is RMS surface error at spatial frequency $n$, and $\alpha_{n,0}$ is the RMS surface error at spatial frequency $n=\lambda/\lambda_0$. 

\begin{figure}
    \centering
    \includegraphics[width=\linewidth]{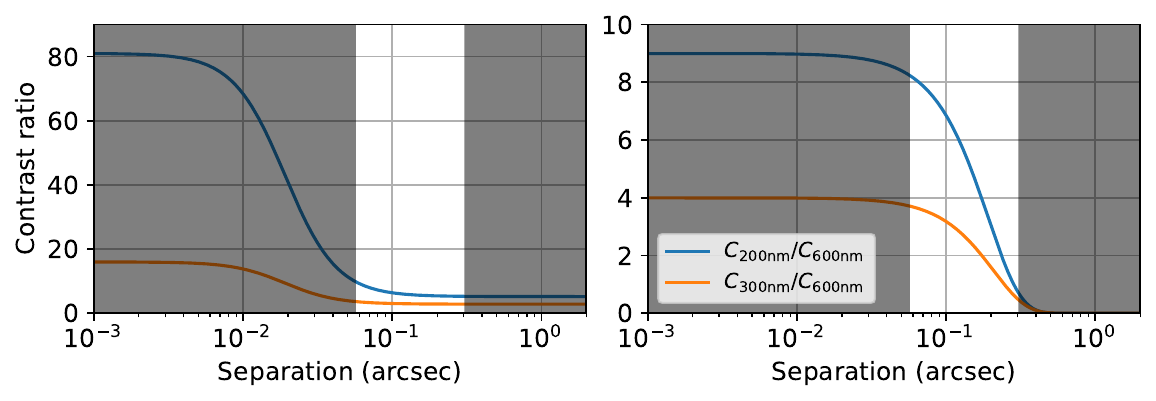}
    \caption{Left: Ratio of contrast at 200 nm (blue) and 300nm (orange) to 600nm for first-order chromatic Talbot residuals, scattering, and beamwalk. A PSD with a knee frequency $b$ at 3 cycles/aperture and exponent $c=2.5$ is assumed. The unshaded region indicates a $9-48\lambda/D$ dark zone at $\lambda=200$nm for a 6.5m aperture (both plots). With these assumptions, the contrast in the dark hole scales roughly as $\lambda^{-1.5}$. Right: Ratio of contrast at 200nm (blue) and 300nm (orange) to 600nm for the DM quantization and noise terms. DM parameters similar to a hypothetical 96x96 MEMS DM are chosen.}
    \label{fig:cratio_both}
\end{figure}

Optical surface and reflectivity errors are commonly specified in terms of a power spectral density (PSD). A PSD representative of optics fabricated for high contrast imaging applications is\cite{mendillo_tolerances, krist_cgi}
\begin{equation}\label{eqn:psd}
    \mathrm{PSD}(n) = \frac{a}{1 + (n/b)^{c}},
\end{equation}
where $a$ sets the peak power of the PSD, $b$ is the knee frequency (in units of cycles per aperture), and and $c$ is an exponent that defines a simple power-law behavior for spatial frequencies beyond the knee. Plugging Equation \ref{eqn:psd} into Equation \ref{eqn:contrast_scaling_general} and evaluating at the limits of small and large separations yields two scaling laws\cite{vangorkom_whitepaper_2025}:
\begin{align}
    C_\mathrm{n\rightarrow 0} &\propto \left(\frac{\lambda_0}{\lambda}\right)^{4} \\
    C_\mathrm{n\rightarrow \infty} &\propto \left(\frac{\lambda_0}{\lambda}\right)^{4-c}
\end{align}

The full form of this equation is plotted in Figure \ref{fig:cratio_both} (left) with illustrative values. A similar result for the scaling of the contrast floor set by DM quantization and actuator noise is given in Equation 6 of Van Gorkom et al.~2025\cite{vangorkom_2025} and plotted in Figure \ref{fig:cratio_both} (right) for a case resembling a high-order MEMS deformable mirror.

% \begin{figure}
%     \centering
%     \includegraphics[width=0.7\linewidth]{cratio_talbot.pdf}
%     \caption{Ratio of contrast at 200nm (blue) and 300nm (orange) to 600nm for first-order chromatic Talbot residuals, scattering, and beamwalk. A PSD with a knee frequency $b$ at 3 cycles/aperture and exponent $c=2.5$ is assumed. The unshaded region indicates a $9-48\lambda/D$ dark zone at $\lambda=200$nm for a 6.5m aperture. With these assumptions, the contrast in the dark hole scales roughly as $\lambda^{-1.5}$.}
%     \label{fig:talbot}
% \end{figure}

% \begin{figure}
%     \centering
%     \includegraphics[width=0.7\linewidth]{cratio_dm.pdf}
%     \caption{Ratio of contrast at 200nm (blue) and 300nm (orange) to 600nm for the DM quantization and noise terms. DM parameters similar to a hypothetical 96x96 MEMS DM are chosen. The unshaded region indicates a $9-48\lambda/D$ dark zone at $\lambda=200$nm for a 6.5m aperture.}
%     \label{fig:dm}
% \end{figure}

We note that the analysis given here is non-exhaustive and primarily illustrative. Other contrast terms (e.g., high-order chromatic Talbot residuals, chromatic residuals from the correction of amplitude errors, and polarization aberrations) show different scaling with wavelength not captured here. Scaling laws for these others terms and the regimes in which individual terms dominate the UV contrast budget will be explored in a future paper.

\begin{figure}
    \centering
    \includegraphics[width=\linewidth]{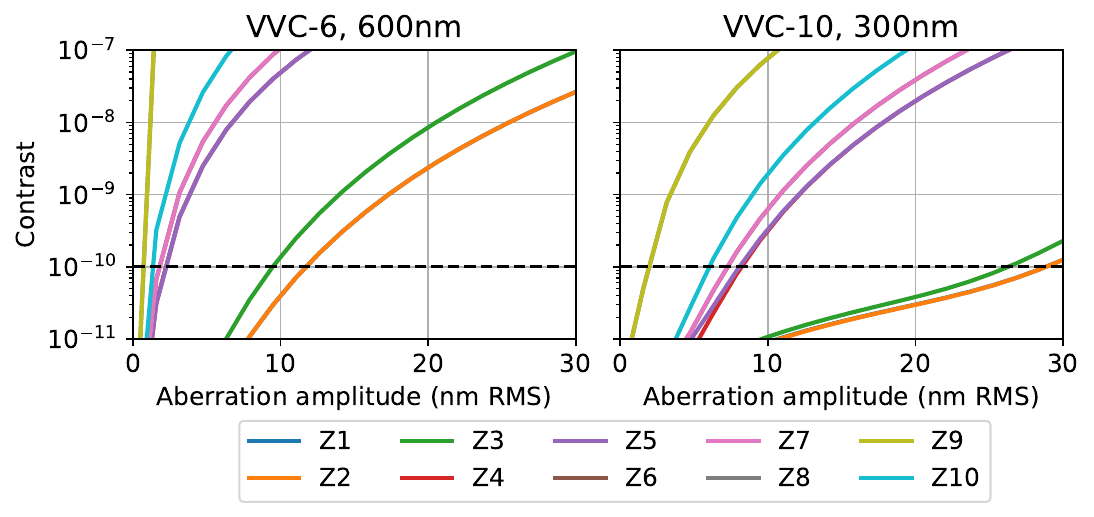}
    \caption{Contrast sensitivity to low-order Zernike aberrations for two coronagraph masks: (left) a charge-6 vector vortex coronagraph (VVC) operating at 600nm and (right) a charge-10 VVC operating at 300nm. Both coronagraph masks achieve the same on-sky IWA, but the latter shows increased robustness to low-order aberrations. Contrast is reported at the IWA. We note that a charge-10 VVC may not be an optimal choice for a large-IWA coronagraph but is simulated here to demonstrate the potential gain from a relaxed NUV IWA.}
    \label{fig:vvc}
\end{figure}

A second argument for the feasibility of a NUV coronagraph from the relaxed NUV IWA requirement is shown in Figure \ref{fig:vvc}. Coronagraph masks tend to trade IWA with sensitivity to low-order aberrations. By exploiting this trade space, it may be possible to design a large-IWA NUV coronagraph that provides the same angular IWA as its visible counterpart while achieving equivalent or greater robustness to low-order aberrations. In addition, large-IWA coronagraphs can be realized with relatively simple coronagraph designs (e.g., pupil-plane apodization masks alone), which may be able to reduce the total number of reflections and address NUV throughput concerns\cite{ctr_2024}.

\section{ONGOING WORK}

The goals of this SAT-funded effort are given in detail in the milestone white paper\cite{vangorkom_whitepaper_2025} and summarized here:
\begin{enumerate}
    \item Development of numerical tools to predict and update UV performance expectations for the SCoOB testbed and a UV coronagraph for HWO.
    \item Component-level metrology of the primary optics, including high-resolution surface measurements, scattered light measurements, and Mueller matrix spectropolarimetry.
    \item Demonstration of UV high contrast imaging on the SCoOB testbed in vacuum. Two complementary coronagraph designs are baselined for this effort---a VVC\cite{mawet_annular_2005,foo_optical_2005,mawet_optical_2009,ruane_vortex_2018,Serabyn19,serabyn_uv_vvc} and a shaped pupil coronagraph (SPC)\cite{kasdin2003,vanderbei2003a,vanderbei2003b}. Testbed results will be compared to simulations to validate the modeling effort and aid in the integration of component-level metrology into the models.
    \item Predictions of science yield informed by  expected UV contrast limits and UV spectral coverage.
\end{enumerate}

\begin{figure}
    \centering
    \includegraphics[width=0.8\linewidth]{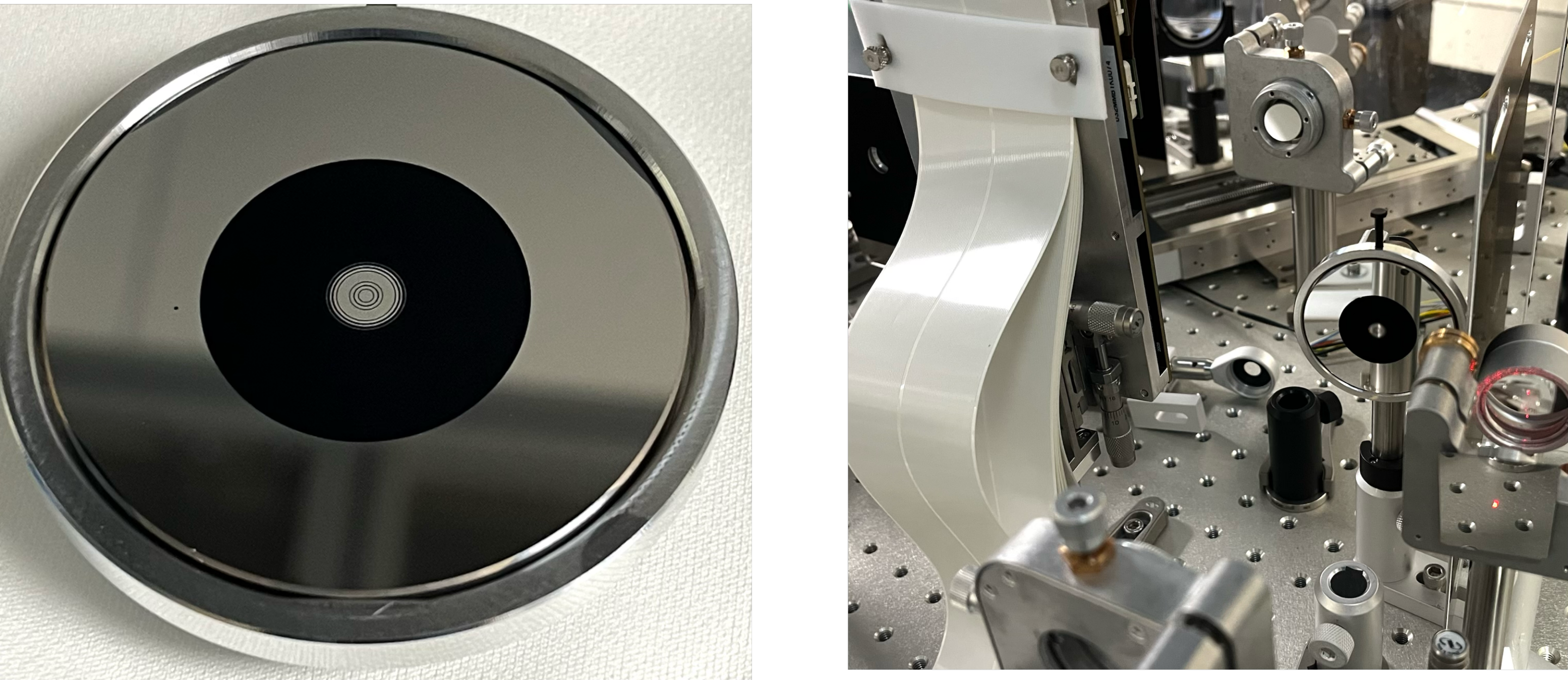}
    \caption{Carbon nanotube SPC mask on a silicon substrate.}
    \label{fig:spc}
\end{figure}

Initial performance predictions of SCoOB\cite{vangorkom_2025} suggest a testbed contrast limit of $3\times10^{-9}$ contrast in a narrow 2\% bandwidth centered at \SI{300}{\nano \meter} and $\lessapprox10^{-8}$ contrast up to a 5\% bandwidth, dominated primarily by chromatic EFC residuals. Preliminary surface roughness measurements with a non-contact atomic force microscope (AFM) capable of achieving ${\ll} \lambda$ resolution have been taken and will be fed into an updated scattering model. A prototype SPC mask targeting a $6-14\lambda/D$ dark hole has been fabricated with carbon nanotubes by Advanced Photonics\cite{hagopian}, and testing in SCoOB to establish visible-wavelength performance with the mask is underway (see Figure \ref{fig:spc}). We expect to begin procurement of a NUV-capable VVC\cite{serabyn_uv_vvc}, NUV sources, and a NUV-capable Mueller matrix microscope for direct retardance measurements in the near future.

\acknowledgements

Portions of this work were supported by the National Aeronautics and Space Administration under Grant 80NSSC25K7458 issued through the Strategic Astrophysics Technology (SAT) program (PI, Van Gorkom), by funding from the Technology Research Initiative Fund (TRIF) of the Arizona Board of Regents, and by generous anonymous philanthropic donations to the Steward Observatory of the College of Science at the University of Arizona.

% References
\bibliography{report} % bibliography data in report.bib

\begin{thebibliography}{10}

\bibitem{tuttle2024_hwo}
Tuttle, S., Matsumura, M., Ardila, D.~R., Chen, P., Davis, M., Ertley, C.,
  Farr, E., Fleming, B., France, K., Froning, C., Grisé, F., Hamden, E.,
  Hennessy, J., Hoadley, K., McCandliss, S.~R., Miles, D.~M., Nikzad, S.,
  Quijada, M., Ravi, I., de~Marcos, L.~R., Scowen, P., Siegmund, O., Vargas,
  C.~J., Vorobiev, D., and Witt, E.~M., ``Ultraviolet technology to prepare for
  the habitable worlds observatory,'' (2024).

\bibitem{ctr_2024}
Chen, P., Pueyo, L.~A., and Siegler, N., ``{A coronagraph technology roadmap
  for future space observatories to directly image Earth-like exoplanets},'' in
  [{\em Space Telescopes and Instrumentation 2024: Optical, Infrared, and
  Millimeter Wave}{\nolinebreak\hspace{0.1em}]},  Coyle, L.~E., Matsuura, S.,
  and Perrin, M.~D., eds.,  {\bf 13092},  130921J, International Society for
  Optics and Photonics, SPIE (2024).

\bibitem{vangorkom_2025}
{Van Gorkom}, K.~J., Anche, R.~M., Mendillo, C.~B., Gersh-Range, J.~A., Hom,
  J., Robinson, T.~D., N'Diaye, M., Lewis, N.~K., Macintosh, B.~A., and
  Douglas, E.~S., ``{Performance predictions and contrast limits for an
  ultraviolet high-contrast imaging testbed},'' {\em Journal of Astronomical
  Telescopes, Instruments, and Systems}~{\bf 11}(4),  042203 (2025).

\bibitem{efc}
Give'on, A., Kern, B., Shaklan, S., Moody, D.~C., and Pueyo, L., ``Broadband
  wavefront correction algorithm for high-contrast imaging systems,''  66910A
  (Sept. 2007).

\bibitem{vangorkom_whitepaper_2025}
{Van Gorkom}, K.~J., Anche, R.~M., Mendillo, C.~B., Gersh-Range, J.~A., Hom,
  J., Robinson, T.~D., N'Diaye, M., Lewis, N.~K., Macintosh, B.~A., and
  Douglas, E.~S., ``{Strategic Astrophysics Technology Technology Milestone
  White Paper: Technology development in UV coronagraphy to enable
  characterization of Earth-like exoplanets},'' (2025).

\bibitem{mendillo_tolerances}
Mendillo, C.~B., Howe, G.~A., Hewawasam, K., Martel, J., Finn, S.~C., Cook,
  T.~A., and Chakrabarti, S., ``{Optical tolerances for the PICTURE-C mission:
  error budget for electric field conjugation, beam walk, surface scatter, and
  polarization aberration},'' in [{\em Techniques and Instrumentation for
  Detection of Exoplanets VIII}{\nolinebreak\hspace{0.1em}]},  Shaklan, S.,
  ed.,  {\bf 10400},  1040010, International Society for Optics and Photonics,
  SPIE (2017).

\bibitem{krist_cgi}
Krist, J.~E., Steeves, J.~B., Dube, B.~D., Riggs, A.~E., Kern, B.~D., Marx,
  D.~S., Cady, E.~J., Zhou, H., Poberezhskiy, I.~Y., Baker, C.~W., McGuire,
  J.~P., Nemati, B., Kuan, G.~M., Mennesson, B., Trauger, J.~T., Saini, N.~S.,
  and Rafels, S.~H., ``{End-to-end numerical modeling of the Roman Space
  Telescope coronagraph},'' {\em Journal of Astronomical Telescopes,
  Instruments, and Systems}~{\bf 9}(4),  045002 (2023).

\bibitem{mawet_annular_2005}
Mawet, D., Riaud, P., Absil, O., and Surdej, J., ``Annular {Groove} {Phase}
  {Mask} {Coronagraph},'' {\em The Astrophysical Journal}~{\bf 633},  1191
  (Nov. 2005).
\newblock Publisher: IOP Publishing.

\bibitem{foo_optical_2005}
Foo, G., Palacios, D.~M., and Swartzlander, G.~A., ``Optical vortex
  coronagraph,'' {\em Optics Letters}~{\bf 30},  3308--3310 (Dec. 2005).
\newblock Publisher: Optica Publishing Group.

\bibitem{mawet_optical_2009}
Mawet, D., Serabyn, E., Liewer, K., Hanot, C., McEldowney, S., Shemo, D., and
  O'Brien, N., ``Optical {Vectorial} {Vortex} {Coronagraphs} using {Liquid}
  {Crystal} {Polymers}: theory, manufacturing and laboratory demonstration,''
  {\em Optics Express}~{\bf 17},  1902 (Feb. 2009).

\bibitem{ruane_vortex_2018}
Ruane, G., Mawet, D., Mennesson, B., Jewell, J.~B., and Shaklan, S.~B.,
  ``Vortex coronagraphs for the {Habitable} {Exoplanet} {Imaging} {Mission}
  concept: theoretical performance and telescope requirements,'' {\em Journal
  of Astronomical Telescopes, Instruments, and Systems}~{\bf 4},  015004 (Mar.
  2018).
\newblock Publisher: SPIE.

\bibitem{Serabyn19}
Serabyn, E., Prada, C.~M., Chen, P., and Mawet, D., ``Vector vortex
  coronagraphy for exoplanet detection with spatially variant diffractive
  waveplates,'' {\em J. Opt. Soc. Am. B}~{\bf 36},  D13--D19 (May 2019).

\bibitem{serabyn_uv_vvc}
Serabyn, E., Ruane, G.~J., and Tabiryan, N.~V., ``{Liquid crystal polymer
  optical vortex phase masks for ultraviolet wavelengths},'' {\em Journal of
  Astronomical Telescopes, Instruments, and Systems}~{\bf 11}(4),  042210
  (2025).

\bibitem{kasdin2003}
Kasdin, N.~J., Vanderbei, R.~J., Spergel, D.~N., and Littman, M.~G.,
  ``Extrasolar {Planet} {Finding} via {Optimal} {Apodized}-{Pupil} and
  {Shaped}-{Pupil} {Coronagraphs},'' {\em The Astrophysical Journal}~{\bf 582},
   1147 (Jan. 2003).
\newblock Publisher: IOP Publishing.

\bibitem{vanderbei2003a}
Vanderbei, R.~J., Spergel, D.~N., and Kasdin, N.~J., ``Circularly symmetric
  apodization via star-shaped masks,'' {\em The Astrophysical Journal}~{\bf
  599},  686--694 (2003).

\bibitem{vanderbei2003b}
Vanderbei, R.~J., Spergel, D.~N., and Kasdin, N.~J., ``Spiderweb masks for
  high-contrast imaging,'' {\em The Astrophysical Journal}~{\bf 590},  593--603
  (2003).

\bibitem{hagopian}
Hagopian, J.~G., Getty, S.~A., Quijada, M., Tveekrem, J., Shiri, R., Roman, P.,
  Butler, J., Georgiev, G., Livas, J., Hunt, C., Maldonado, A., Talapatra, S.,
  Zhang, X., Papadakis, S.~J., Monica, A.~H., and Deglau, D., ``{Multiwalled
  carbon nanotubes for stray light suppression in space flight instruments},''
  in [{\em Carbon Nanotubes, Graphene, and Associated Devices
  III}{\nolinebreak\hspace{0.1em}]},  Pribat, D., Lee, Y.-H., and Razeghi, M.,
  eds.,  {\bf 7761},  77610F, International Society for Optics and Photonics,
  SPIE (2010).

\end{thebibliography}
\bibliographystyle{spiebib} % makes bibtex use spiebib.bst

\end{document}